\documentclass[12pt]{iopart}
\usepackage{iopams} 
\usepackage{graphicx}
\usepackage{subfigure}
\usepackage{placeins}
\usepackage{url}
\usepackage{multirow}
\begin{document}
	
	\title{A laser interferometer accelerometer for vibration sensitive cryogenic experiments}
	
	\author{R.Bajpai$^1$, T.Tomaru$^2$, K.Yamamoto$^3$, T.Ushiba$^4$, N.Kimura$^4$, T.Suzuki$^5$, T.Yamada$^6$, T. Honda$^6$}
	
\address{$^1$ The Graduate University for Advanced Studies, Department of Accelerator Science, School of High Energy Accelerator Science, High Energy Accelerator Research Organization (KEK) Tsukuba, Ibaraki, 305-0801, Japan}

\address{$^2$ National Astronomical Observatory of Japan, 2 Chome-21-1 Osawa, Mitaka, Tokyo, 181-8588, Japan}

\address{$^3$ University of Toyama, 3190 Gofuku, Toyama, 930-8555, Japan}

\address{$^4$ The University of Tokyo Institute for Cosmic Ray Research Kamioka Observatory, Higashimozumi 238, Kamioka, Hida, Gifu, 506-1205, Japan}

\address{$^5$ The University of Tokyo Institute for Cosmic Ray Research, Kashiwanoha 5-1-5, Kashiwa, Chiba, 277-8582, Japan}

\address{$^6$ High Energy Accelerator Research Organisation, 1-1 Oho, Tsukuba, Ibaraki, 305-0801, Japan}

\ead{bajpai@post.kek.jp} 
\begin{abstract}
	 Monitoring motion originating from ultra low-temperature cooling systems like cryocoolers is important for vibration sensitive cryogenic experiments like KAGRA. 
	 Since no commercial cryogenic accelerometers are available, we developed a compact self-calibrating accelerometer with a Michelson interferometer readout for cryogenic use. 
	 Change in calibration factor and drop in interferometer output originating from temperature drop were the main concerns which were tackled. 
	 Sensitivity of $3.38\times10^{-11}$ m/$\sqrt{\mathrm{Hz}}$ at 1 Hz was achieved at 300 K. 
	 The accelerometer was tested inside the KAGRA cryostat; showed stable operation down to 12 K in 0.1-100 Hz band with only 1\% visibility drop. 
	 Our accelerometer can be employed in low vibration cryogenic environment for a multitude of applications.
\end{abstract}



\section{Introduction}
Cryogenic temperatures are needed in various high precision experiments where thermal fluctuations can be a noise source. 
Mechanical cryocoolers provide a safe, low cost and reliable way to achieve cryogenic temperature.
However, the mechanical vibrations originating from the operation of cryocoolers are a noise source in vibration sensitive experiments like KAGRA, ILC and CUROE \cite{1,2,3}. 
Therefore, considerable efforts have been made to develop low vibration pulse tube cryocoolers \cite{4}. 
But the development of high sensitivity cryogenic accelerometers to monitor these vibrations and design active damping systems is still a key task as no such devices are commercially available.

Large-scale Cryogenic Gravitational wave Telescope (KAGRA), is a second-generation gravitational-wave detector (GWD) located in Kamioka mine, Japan. 
The detector is under commissioning and will join Advanced LIGO \cite{5}, and Advanced VIRGO \cite{6} for a joint O4 observation run in 2022. 
The features that distinguish KAGRA from other GWDs are its underground location and cryogenic operation of the four main mirrors. 
The underground location provides a quiet site with low seismic and gravity gradient noise \cite{7,8}, while the cryogenic operation cools the mirrors down to 20 K, reducing the thermal noises. 
However, vibration originating from the cryocooler operation and structural resonances of the cryostat can contaminate the detector sensitivity making the quite underground location redundant. 
Therefore, monitoring and characterization of the vibration inside cryostat are critical for the optimum noise performance of KAGRA. 

For the aforementioned reason, we developed a sensitive cryogenic accelerometer and evaluated its performance using the low vibration environment of KAGRA cryostat.
\section{Requirements} \label{sec:2}

\begin{figure} [t]
	\centering
	\includegraphics[width=0.9\textwidth]{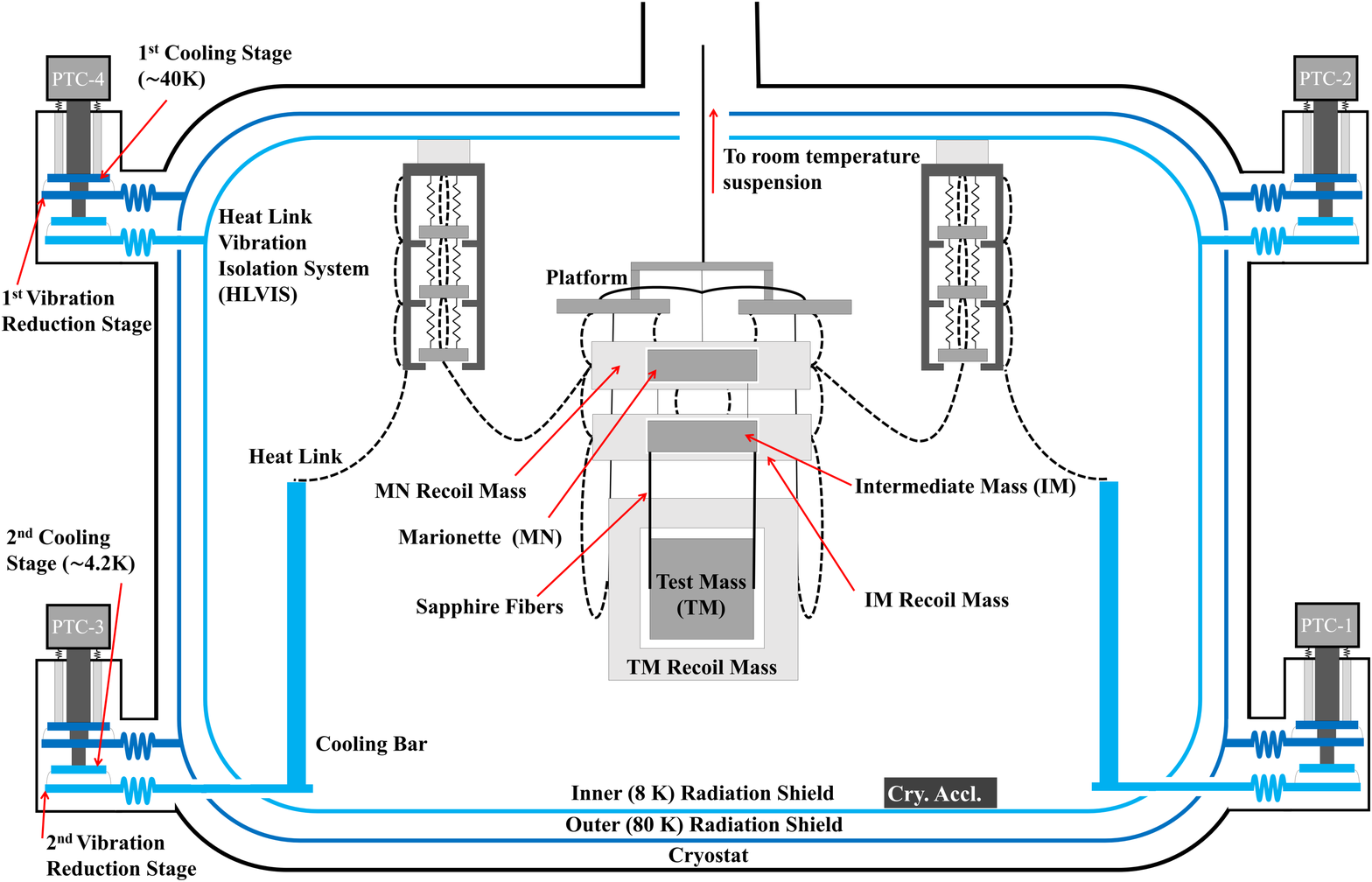}
	\caption{Schematic layout of KAGRA cryogenic system. The outer black block is the vacuum chamber; the inner boxes are 80 K and 8 K radiation shield; PTC, CB, HL, HLVIS, are cryocoolers, heat conductor (cooling bar), heat-links, heat-links vibration isolation system, respectively and form the cooling system. The cryogenic payload consists of the platform (PF), marionette (MN), intermediate-mass (IM), test mass (TM) and their recoil masses MNR, IRM, RM. The tiny grey rectangle shows the cryogenic accelerometer mounting position during its testing.}
	\label{fig:1}
\end{figure}

The specifications for the cryogenic accelerometer were set based on its application for KAGRA cryostat vibration analysis. 
In this section, we briefly describe the KAGRA cooling system and then specify the requirements for the cryogenic accelerometer. 
\Fref{fig:1} shows the layout of KAGRA cryogenic system. 
The main test masses (23 kg sapphire mirrors) in KAGRA are suspended from a nine-stage suspension; the bottom four stages, including the mirror, are called 'Cryogenic Payload' \cite{9}. 
It is cooled within a double-radiation shield cryostat using four ultra-low vibration pulse-tube cryocoolers \cite{10}. 
The 1st stage of each cryocooler is connected to the outer (80 K) shield, while 2nd stage of two cryocoolers is connected to the inner (8 K) shield. 
Thermal radiation from shields effectively cools the payload down to $\sim 100$ K. 
In addition, 2nd stage of the other two cryocoolers are connected to 6N (6 Nine, $99.9999\%$ purity) aluminium  cooling bars that are in thermal contact with the payload through thin 6N Al heat-links \cite{11} to facilitate conduction cooling of the test mass down to 20 K. 
Cryocooler vibration can couple to test mass through these heat-links. 

The requirements for the cryogenic accelerometer are:

\begin{enumerate}
	\item \textit{Calibration}: The accelerometer needs to be self-calibrating because the calibration factor will change with temperature, and there is no reference accelerometer available to re-calibrate it at low temperature.
	\item \textit{Temperature}: The accelerometer should be able to operate inside KAGRA cryostat, which is typically around 11-15 K when cooling down. 
	\item \textit{Sensitivity}: The underground location at KAGRA  provides a quiet site; typical seismic motion being $3\times10^{-9}$ m/$\surd$Hz at 1 Hz, which is 2 order of magnitude smaller than seismic motion at surface. 
	We set the minimum sensitivity required to be below the KAGRA seismic level. 
	\end{enumerate}
\section{Development and Setup}

Our accelerometer is based on a high sensitivity seismometer developed for seismic noise measurement at Kamioka site (the relative displacement between a pendulum and its suspension point is measured with a Michelson Interferometer) \cite{12,13}. 
In this section, we describe the accelerometer layout and design of individual components.

\subsection{Accelerometer Layout}

\begin{figure} 
	\centering
	\subfigure[] {\includegraphics[width=0.4\textwidth]  {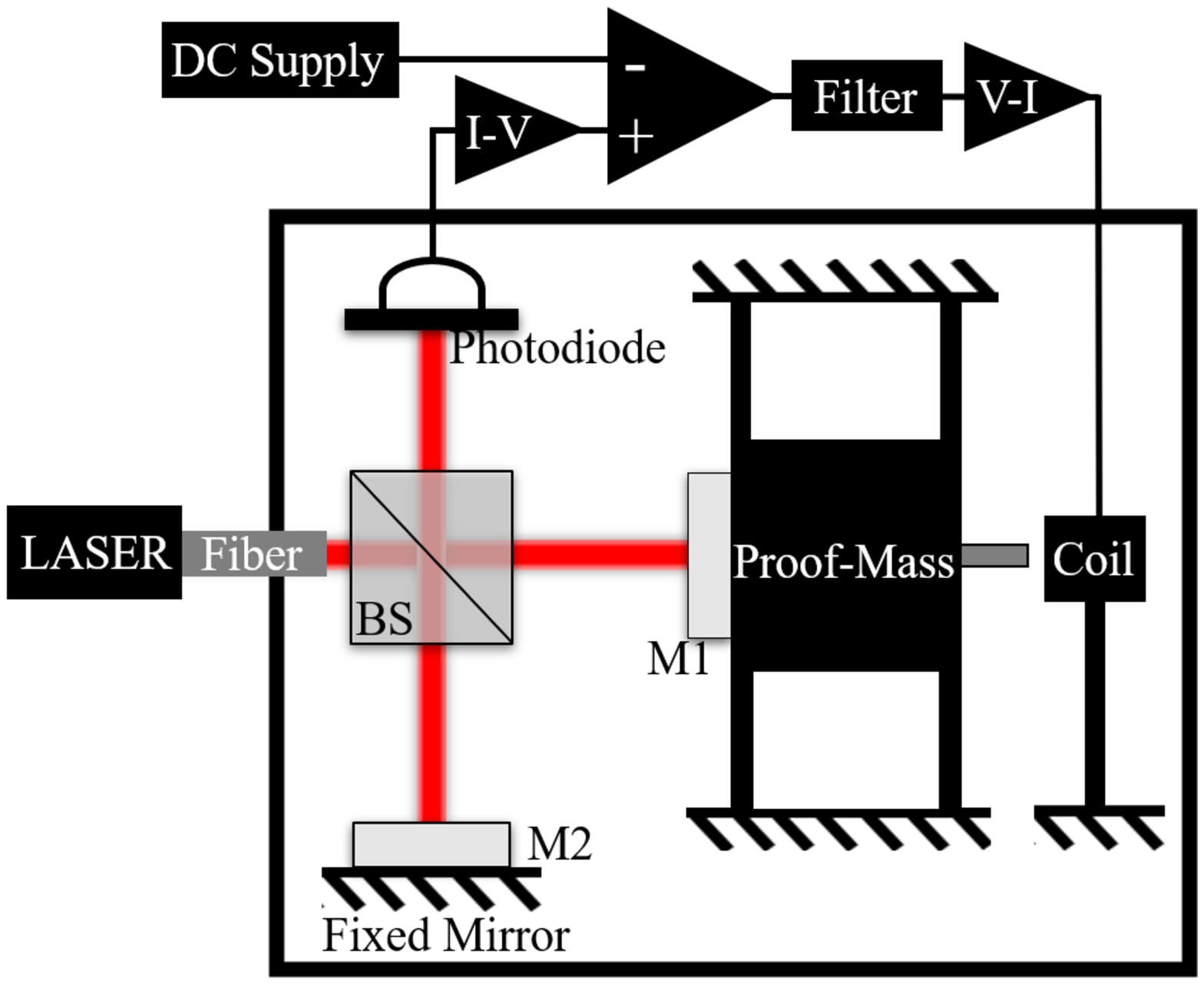}\label{fig:2-A}}	
	\subfigure[]{\includegraphics[width=0.5\textwidth]{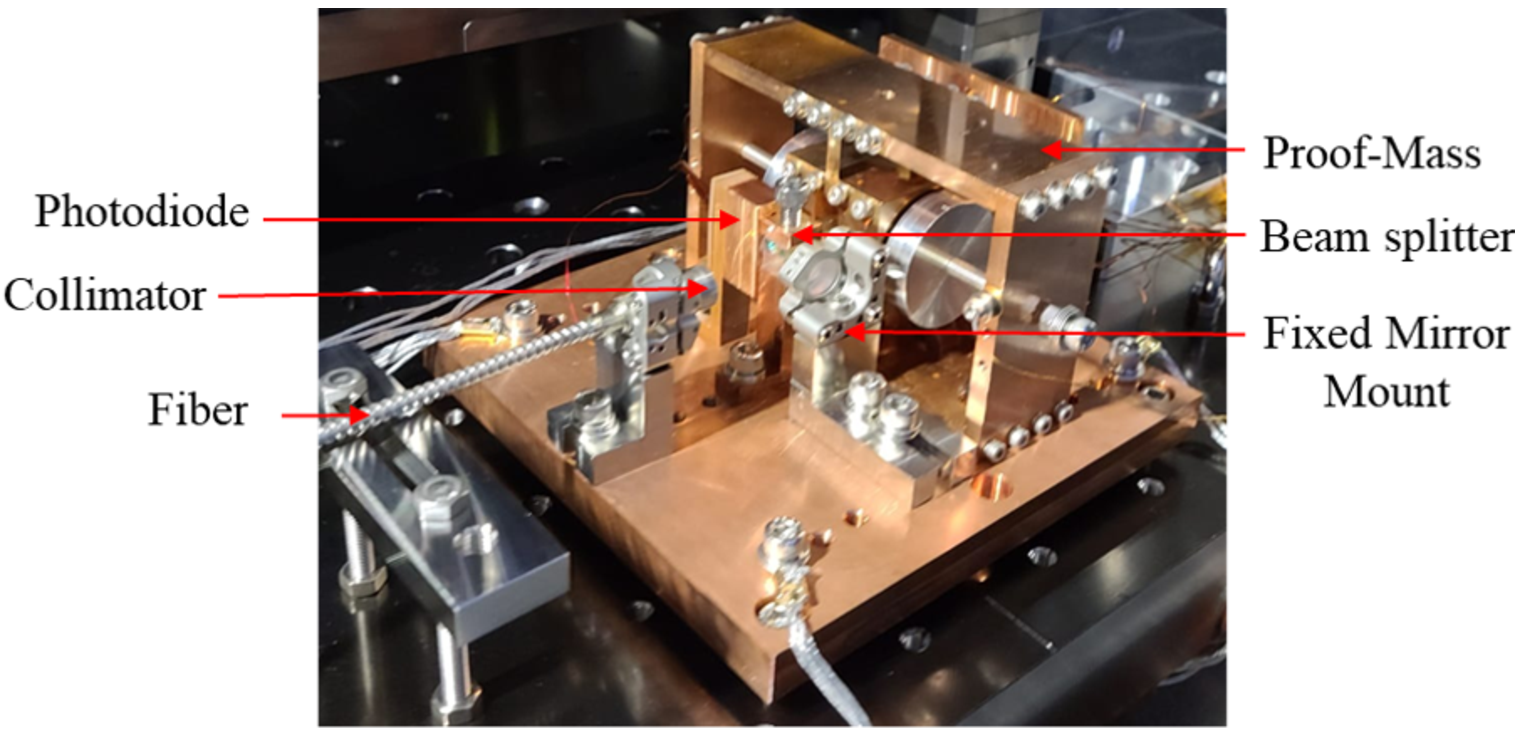}\label{fig:2-B}}
	\subfigure[]{\includegraphics[width=0.7\textwidth]{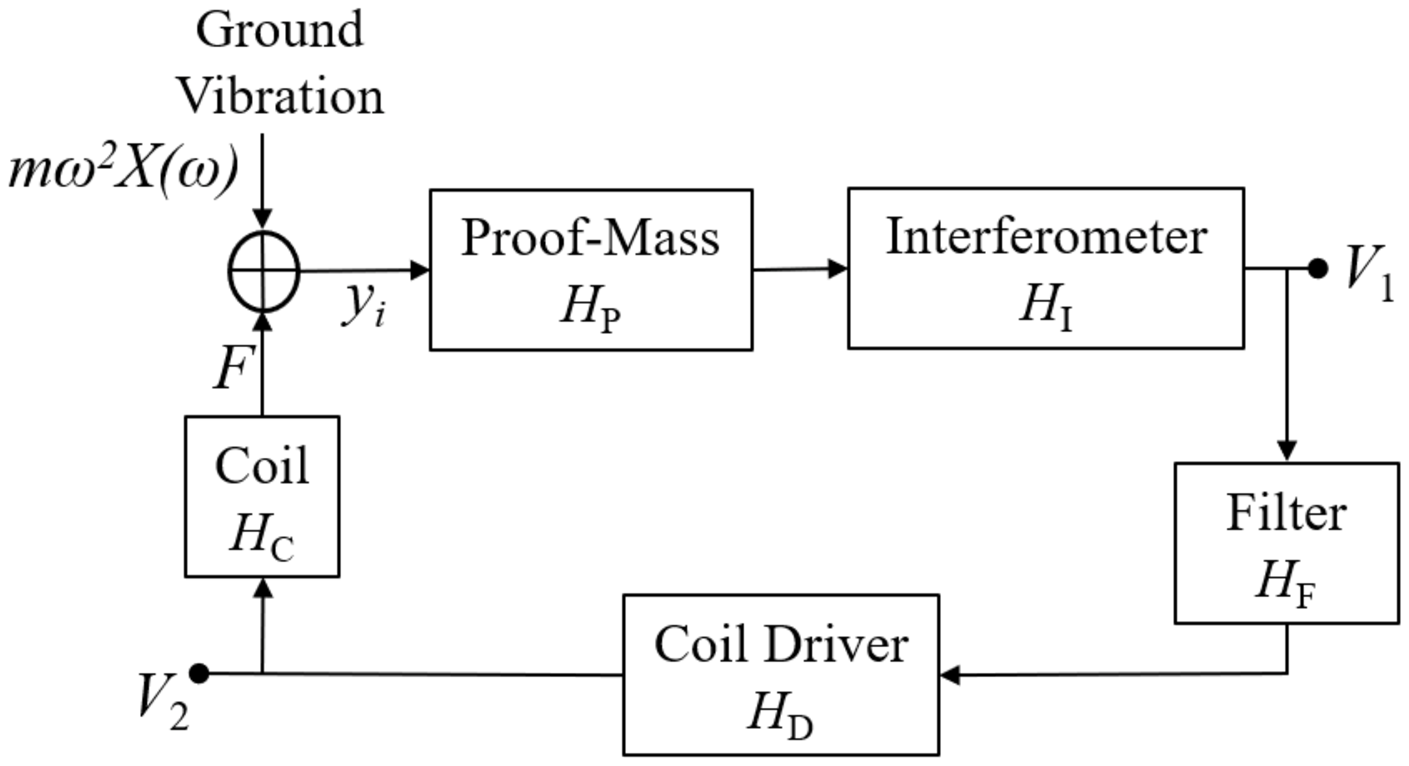}\label{fig:2-C}} 
	\caption{Cryogenic Accelerometer: (a) Schematic layout of cryogenic accelerometer. The LASER is introduced through an optical fiber. The black square denotes the components (proof-mass, fixed mirror, photo-diode, beamsplitter and coil) which are housed inside the cryostat. (b) Photo of cryogenic accelerometer in KAGRA cryostat. The laser diode and circuit box are housed outside the cryostat. (c) Block diagram of cryogenic accelerometer in locked state.}
\end{figure}

Based on our requirements, we designed an accelerometer with a Michelson Interferometer (MI) readout that operates in a closed-loop. 
\Fref{fig:2-A} shows the schematic layout of the cryogenic accelerometer. 
The MI is formed by two mirrors (M1 and M2), M1 glued to a proof mass and M2 fixed rigidly to the base plate; the laser is introduced into the cryostat through an optical fiber. 
The interference signal, readout by a photo-diode, detects the relative displacement between ground and proof mass. 
This signal is used as feedback to coil magnet actuators that lock the proof mass in linear range of the interferometer. 
Below unity gain frequency, the force from actuators to proof mass is proportional to the inertial vibration of the base plate ("proof-mass"-"beamsplitter" direction). 
Vibration can be derived from the feedback signal.

\Fref{fig:2-C} shows the block diagram of the cryogenic accelerometer where $H_\mathrm{{P}}, H_\mathrm{{I}}, H_\mathrm{{F}},H_\mathrm{{D}}, H_\mathrm{{C}} $ are proof-mass, interferometer, filter, coil driver and coil transfer functions, respectively; $V_\mathrm{{1}}$ and $V_\mathrm{{2}}$ are the  interferometer output and coil input (feedback signal), respectively. 
The inertial force that acts on the proof-mass ($m$) for some ground vibration ($X(\omega)$) is $m \omega^{2}X(\omega)$, and the feedback force from the actuator to lock the mass is $\textit{F}$. 
From the block diagram, the expression of vibration can be derived as:
\begin{eqnarray}\label{eq1}
	\centering
	|X(\omega)|=\frac{1}{m\omega^{2}}\left| \frac{1+G}{G}\right|  |H_\mathrm{{C}}| |V_\mathrm{{2}}|
\end{eqnarray}
where
$G=H_\mathrm{{P}} H_\mathrm{{I}} H_\mathrm{{F}} H_\mathrm{{D}} H_\mathrm{{C}}$ is the open-loop transfer function.

In equation \eref{eq1}, $ |H_\mathrm{{C}}|\left| {1+G}/{G}\right|$ is the calibration factor ($C$) and converts the voltage $V_\mathrm{{2}}$ to the feedback force. 
Dividing the feedback force with proof-mass weight ($m$) gives us the inertial acceleration of the base plate. 
Displacement spectral density is obtained from the acceleration spectrum by dividing by $\omega^{2} [=(2\pi f)^{2}]$. 
Finally, the expression for vibration can be simplified to:
\begin{eqnarray}\label{eq2}
	\centering
	|X(\omega)|=\frac{C}{m\omega^{2}} |V_\mathrm{{2}}|
\end{eqnarray}

\subsection{Cryogenic Compatibility}

Cooling down accelerometer will cause attenuation at interferometer output due to: (a) reduction in PD responsitivity and (b) signal loss from a drop in fringe visibility originating from angular misalignment due to thermal shrink of each component; this can cause the accelerometer to be inoperable at low temperatures; to minimize such loss, individual components were selected and tested.

\subsubsection{Optics}
\begin{figure} [b]
	\centering
	\includegraphics[width=0.9\textwidth]{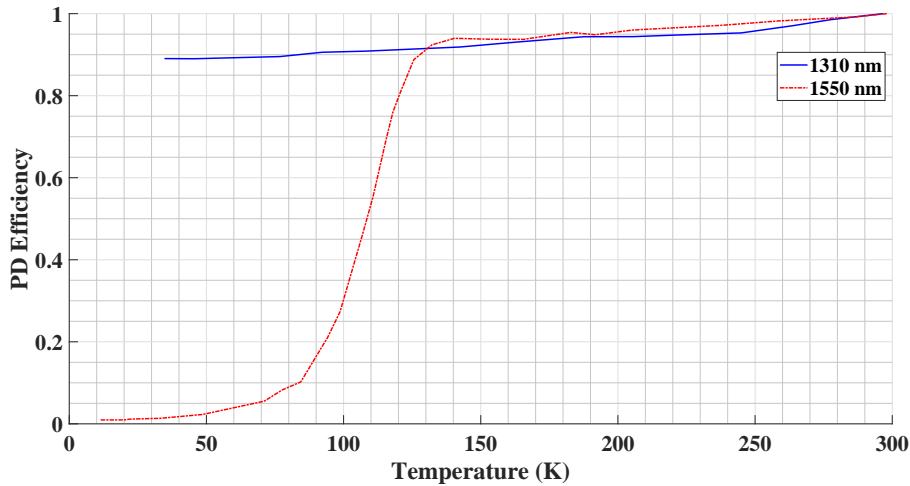}
	\caption{Photo-diode efficiency (FGA21) as a function of Temperature, for 1310 nm (blue line) and 1550 nm (red dashed line) light.}
	\label{fig:3}
\end{figure}

The interference signal was read out using Thorlabs FGA21 photo-diode \cite{14} (active area: $\phi2$ mm), which is known to be compatible at cryogenic temperature through our testing. 
Further cooling tests showed that this cryogenic compatibility is wavelength dependent; the results are shown in \fref{fig:3}. 
Note that for measurement with 1330 nm laser final temperature of PD was 32 K; no drop in efficiency was observed at lower temperatures when the PD was cooled with accelerometer setup in \sref{sec:4.2} and \ref{sec:4.4} down to 16 K and 12 K, respectively.

Based on the photo-diode performance, a 15 mW, 1310 nm LASER diode was used as the light source and introduced into the cryostat through a single-mode fiber coupled to a collimator (beam waist-0.7 mm).
 
A 10 mm beamsplitter (splitting ratio: 47\% (transmitted) to 46.7\% (reflected)) and two ½” fused silica mirrors (reflectivity$\sim 99.5\%$) form the MI with arm-length of 9 mm. 

\subsubsection{Mechanical Scheme}

Mirror misalignment due to thermal shrink of accelerometer components can cause a drop in fringe visibility. 
The mechanical components were designed to minimize the misalignment.
\begin{itemize}
	
	\item \textit{Proof Mass:} The  proof mass is a cube of mass, $m=0.555$ kg made of Oxygen-Free Copper (OFC), which is suspended from an OFC frame by 4, Beryllium Copper (BeCu) blade springs of length, $l= 15$ mm and thickness, $t=100\ \mu$m, which are shown in \fref{fig:4}. 
	The resonant frequency is 9.5 Hz. 
	A much lower resonant frequency can be achieved using an inverted pendulum of the same size; however, this rigid proof-mass design reduces the impact of radiation shield tilt on fringe visibility as it restricts the motion of proof-mass in the direction parallel to the mounting surface. 
	Also, the proof mass makes it possible to extend the scheme in vertical direction measurement with minor changes.
	
	\begin{figure} 
		\centering
		\includegraphics[width=0.35\textwidth]{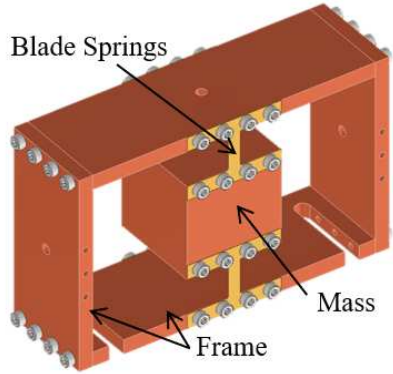}
		\caption{Proof mass for cryogenic accelerometer. The mass is a cube of side 0.4 m made of Oxygen-Free Copper (OFC). It is suspended from an OFC frame by 4, Beryllium Copper (BeCu) blade springs of length, $l= 15$ mm and thickness, $t=100\ \mu$m.}
		\label{fig:4}
	\end{figure}
	
	\item \textit{Fixed Mirror Mount:} A large contribution to visibility drop is expected to come from the shrink of fixed mirror mount; \textit{Thorlabs, 'POLARIS-K05} mount \cite{15} (shown in \fref{fig:2-B}) was selected as our cooling test showed it to have a small angular drift of approximately $0.04^{\circ}$ and $0.05^{\circ}$, along pitch and yaw axis respectively. 
	In this cooling test, the mount with a mirror was cooled down inside a cryostat; a laser reflected off the mirror falls on a position-sensitive detector (PSD). 
	The drift of this reflected beam is monitored as the mount is cooled. 
	\Fref{fig:5} shows the result of mirror mount tilt; drift observed for pitch axis is due to non-uniform cooling of the mount and recovers once a stable temperature of 16 K is achieved. 
	To precisely align the interferometer, fiber-coupled collimator was also mounted on this mount, as shown in \fref{fig:2-B}.
	\vspace{1 mm}
	\begin{figure} [t]
		\centering
		\includegraphics[width=0.9\textwidth]{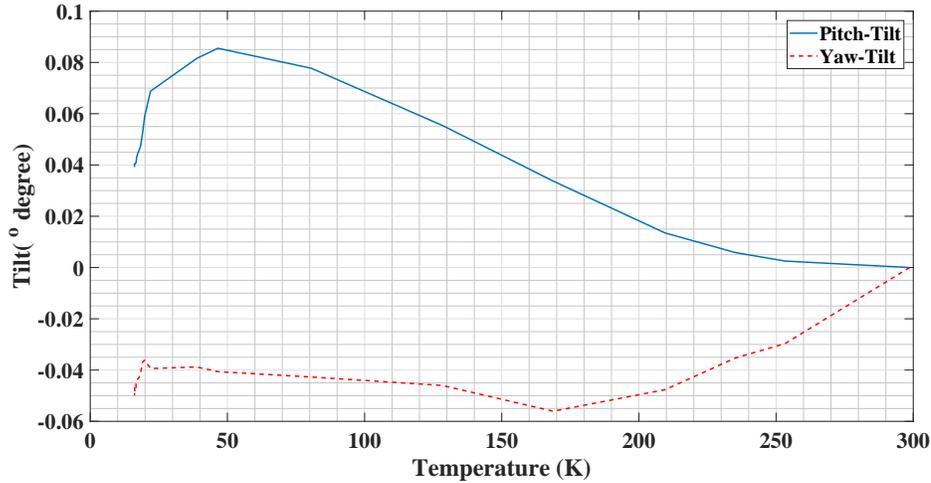}
		\caption{Result of fixed mirror mount cooling test.}
		\label{fig:5}
	\end{figure}
\end{itemize}

The components were mounted rigidly on a base plate of dimensions: $170\times170\times15$ mm$^3$ supported by 3-M6 screws to establish three-point contact. 
With proper alignment, fringe visibility of $98.5\%$ was achieved. 
From the results of PD efficiency and fixed mirror mount drift measurement, we expect the attenuation at interferometer output will not be an issue at cryogenic temperature; this was confirmed with a cooling test as reported in \sref{sec:4.2}.
\\
\subsubsection{Electronics}

Electronics for accelerometer are housed outside the cryostat, as seen in \fref{fig:2-A}. 
A Lake Shore temperature sensor (\textit{DT-670}) is bolted to the base plate to monitor the accelerometer temperature.

The current signal from the PD is converted to voltage signal and passed through a differential amplifier to remove the dc offset, generating the error signal. 
The error signal is fed to the controller circuit (band-pass filter + coil driver), tuning the unity gain frequency of the open-loop transfer function (G) above 100 Hz and generating the actuating current signal. 
This signal drives coil magnet actuators, locking the proof-mass. 
The Samarium Cobalt (SmCo) magnets are fixed to the back of the proof-mass with Epoxy (3M\textsuperscript{TM} Scotch Weld\textsuperscript{TM} \textit{DP-190}). 
SmCo magnets are used as their magnetism doesn't change significantly between the room and cryogenic temperature \cite{16}.

\subsection{Calibration}\label{sec:3.4}

The calibration factor (\textit{C}) will change as the accelerometer is cooled because: (a) actuator efficiency will be lower due to reduction in magnetism of SmCo magnets and change in the coupling factor between the magnet and coil, (b) optical loss at interferometer output from a drop in fringe visibility and (c) reduction in PD responsitivity. 
However, the accelerometer can be calibrated at any temperature by measuring open-loop transfer function ($G$) and coil transfer function ($H_{C}$) as follows:

\begin{itemize}
	
	\item \textit{Open-loop transfer function($G$)} is measured at interferometer output by swept sine excitation.
	
	\item \textit{Coil transfer function ($H_\mathrm{{C}}$)} shown in \fref{fig:2-C} is calculated from coil to interferometer transfer function $(V_\mathrm{{1}}/V_\mathrm{{2}})$ as:
	\begin{eqnarray}\label{eq11}
	H_\mathrm{{C}}=\frac{(V_\mathrm{{1}}/V_\mathrm{{2}})}{H_\mathrm{{I}}H_\mathrm{{P}}} 
	\end{eqnarray}
	where, 	
	$H_\mathrm{{P}}$ is the proof-mass transfer function and is approximated to $1/m(2\pi f_\mathrm{{r}})^{2}$ for frequency, $f\ll f_\mathrm{{r}}$ (resonant frequency).
	
	$H_\mathrm{{I}}$ is the interferometer transfer function.
	
	\item \textit{Interferometer transfer function} can be calculated when there is no fed back to the proof mass (free-swinging Michelson) as: 
	\begin{eqnarray}
	H_\mathrm{I}=\frac{4\pi V_{ \mathrm{1free}}}{\lambda}
	\end{eqnarray}
	where,
	
	$\lambda =1310$ nm is the LASER wavelength
	
	$V_{\mathrm{1free}}$ is the amplitude of interferometer output when the proof mass is free
	
\end{itemize}

\begin{figure} [h!]
	\centering
	\includegraphics[width=1\textwidth]{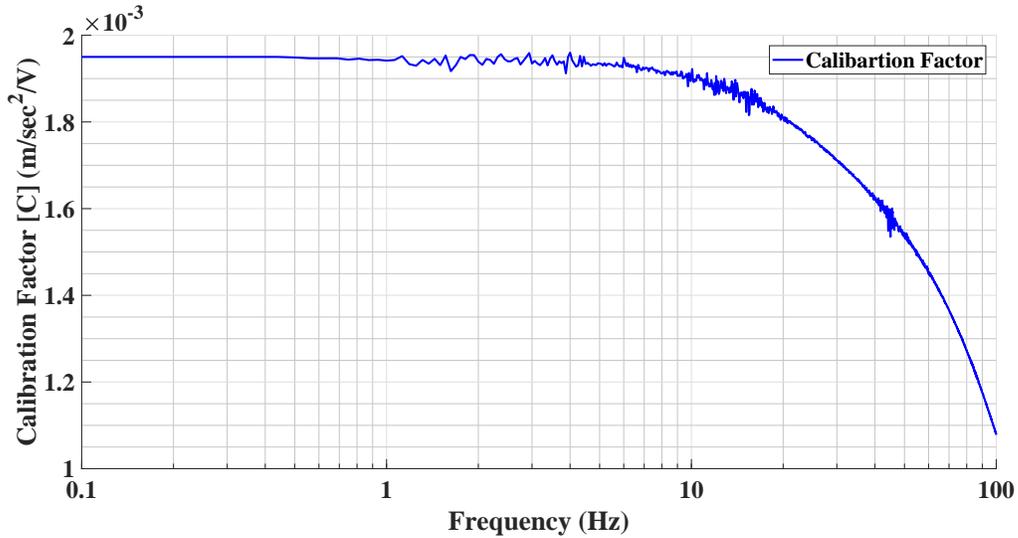}
	\caption{Calibration factor of cryogenic accelerometer determined using wavelength calibration. The interferometer transfer function, $H_\mathrm{I}$, was $16.54$ nm/V, and the coil transfer function $H_\mathrm{C}$ was $0.0019$ N/V.}
	\label{fig:6}
\end{figure}

We call this 'wavelength calibration' and can be used to calibrate the accelerometer at any temperature. 
\Fref{fig:6} shows the calibration factor measured at room temperature and atmospheric pressure.

\section{Testing}
The accelerometer was tested at room and cryogenic temperature to confirm it satisfies the requirements set in \sref{sec:2}. 
In this section we report the results of these tests and accelerometer performance.
\\
\subsection{Performance Comparison} 
To confirm the reliability of wavelength calibration as described in \sref{sec:3.4}, the cryogenic accelerometer and a commercial accelerometer (RION LA-50) were mounted (in air) together on an optical table; vibration was measured simultaneously. 
The spectrum measured by the two accelerometers is consistent, and the coherence between the output of the two is almost 1 over the entire spectrum, as seen in \fref{fig:6}. 
We conclude that the absolute value of our accelerometer output is reliable. 

\begin{figure}[h!]
	\centering
	\includegraphics[width=1.1\textwidth]{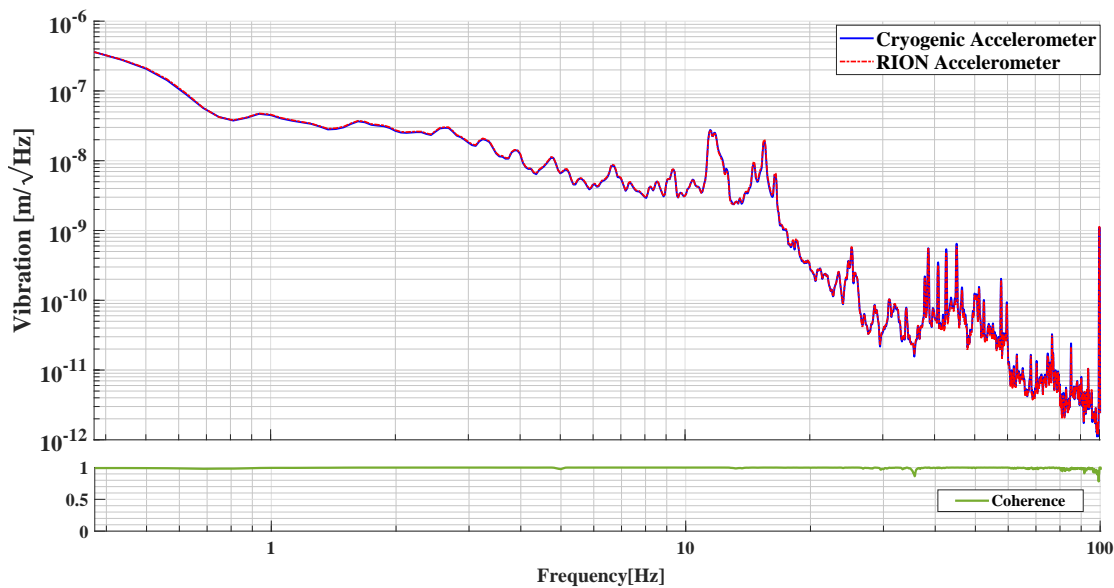}
	\caption{Vibration spectrum of the optical table measured; by our accelerometer (blue) and a commercial accelerometer (red) (top). Coherence (green) between output of two accelerometers (bottom).}
	\label{7}
\end{figure}

\subsection{Cryogenic Testing} \label{sec:4.2}

\begin{figure} [h!]
	\centering
	\subfigure[] {\includegraphics[width=1.1\textwidth]{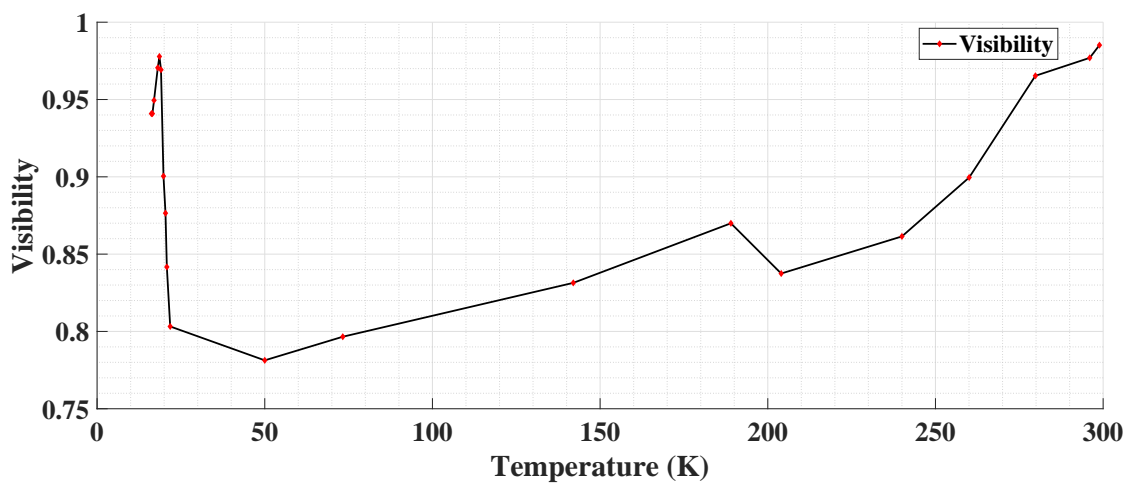}\label{fig:8-A}}	
	\subfigure[]{\includegraphics[width=1.1\textwidth]{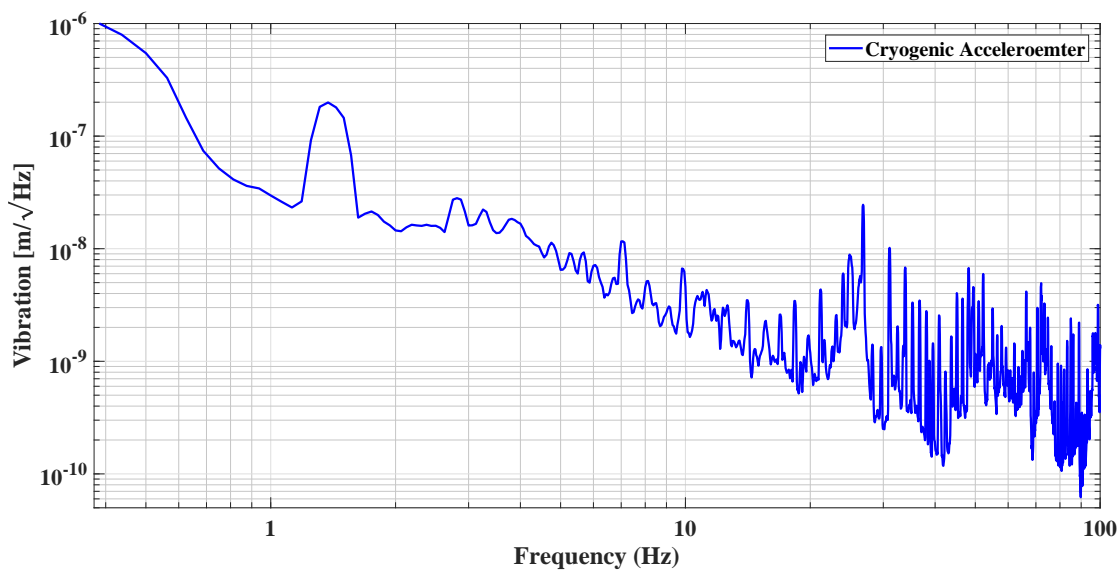}\label{fig:8-B}} 
	\caption{Results of cryogenic testing of accelerometer. (a) Visibility of Michelson Interferometer Vs temperature of accelerometer. (b) Vibration spectrum of a cryostat in KEK at 16 K. The peak at 1.37 Hz and its harmonics are from cryocooler operation.}
\end{figure}

Next, the accelerometer was cooled down in a cryostat in our laboratory at the High Energy Accelerator Research Organisation (KEK) to evaluate the low-temperature performance. 
Initial visibility of accelerometer at 297 K was $98.5\%$; \fref{fig:8-A} shows interferometer visibility  as accelerometer was cooled. 
The intermediate drop is linked to the pitch drift of the fixed mirror mount and non-uniform cooling of individual components. 
Once the accelerometer reached a stable temperature of 16 K, visibility recovered to $94\%$.

Finally, the accelerometer was re-calibrated, and vibration of the cryostat was measured. 
We confirmed that thermal shrink did not cause any significant visibility drop, and accelerometer shows stable operation down to 16 K.

\subsection{Sensitivity}

\begin{figure} [h!]
	\centering
	\includegraphics[width=1.1\textwidth]{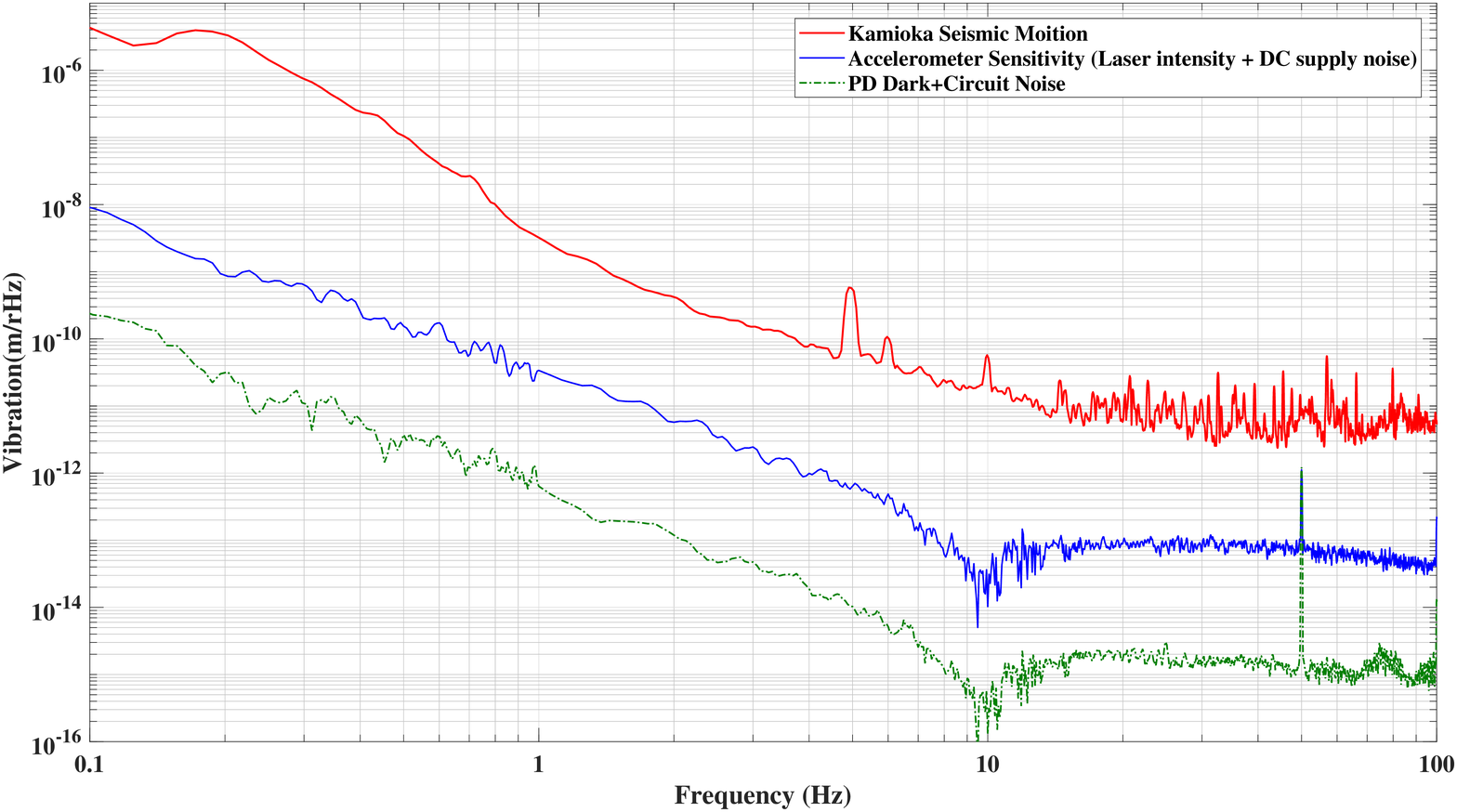}
	\caption{Sensitivity of cryogenic accelerometer dominated with laser intensity and DC supply electronic (blue) compared to PD dark+ circuit noise (green) and Kamiaoka seismic motion (red).}
	\label{fig:9}
\end{figure}

The main noise sources for the accelerometer are the intensity noise of the laser and electronic noise of the circuits, photo-diode and DC supply. 
To evaluate the sensitivity of accelerometer we block the proof-mass arm with a beam dump; under such conditions, there is no interference, and the output shows the approximate noise level ($V_\mathrm{{noise}}$) of the accelerometer itself due to laser intensity and electronic noise of the DC supply.  
The minimum vibration that can be measured by our accelerometer is shown in blue in \fref{fig:9}. 
The PD dark and circuit noise are evaluated by measuring $V_\mathrm{{noise}}$ when the laser and dc supply is turned off.
\Fref{fig:9} shows that the noise level of the cryogenic accelerometer is lower than Kamaioka seismic motion; therefore, it can be tested in KAGRA cryostat in 0.1-100 Hz  bandwidth. 

\subsection{Onsite Testing} \label{sec:4.4}

\begin{figure}[h!]
	\centering
	\includegraphics[width=0.4\textwidth]{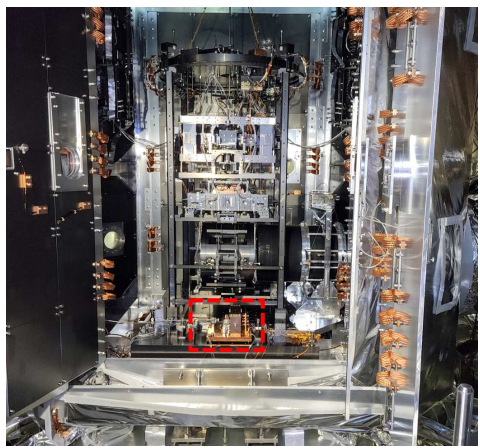}
	\caption{Cryogenic Accelerometer (red box) mounted inside one of the KAGRA cryostat.}
	\label{fig:10}
\end{figure}

\begin{figure}[h!]
	\centering
	\includegraphics[width=1\textwidth]{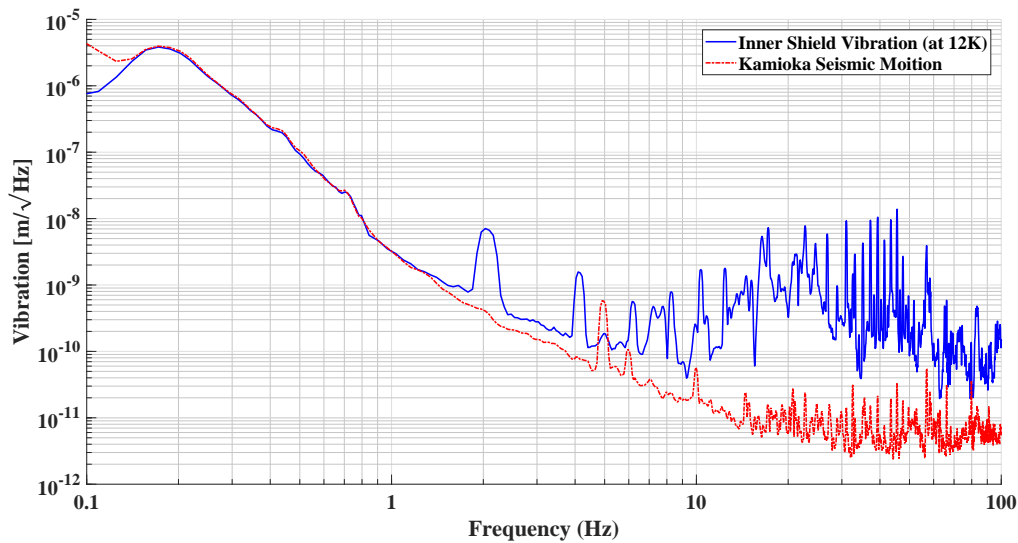}
	\caption{Vibration Spectra of KAGRA cryostat shown in \fref{fig:10}. At the time of measurement inner shield and accelerometer were at 14.7 K and 12 K, respectively, and all cryocoolers were operating. The seismic motion was measured by a commercial accelerometer (RION LA-50) placed next to the cryostat in air.}
	\label{fig:11}
\end{figure}

As the vibration level in KEK is 10-100 times larger than Kamioka, we tested the cryogenic accelerometer in KAGRA cryostat, \fref{fig:10}. 
The final temperature of the accelerometer and inner shield was 12 K and 14.7 K, respectively; the final interferometer visibility was $96$\%. 
\Fref{fig:11} shows the vibration spectrum of inner shield at $12$ K when all cryocoolers are operating. 
We confirmed that our developed accelerometer can operate in low vibration cryogenic environment. 

\section{Conclusion}

We developed a cryogenic accelerometer with a Michelson interferometer readout. The mechanical and optical components were selected, designed and tested to minimize loss of interferometer output at cryogenic temperature. Accelerometer was tested at room and cryogenic temperature; it satisfied the sensitivity, temperature and calibration requirements.  We installed the accelerometer in KAGRA cryostat, and it showed stable operation down to 12 K. We conclude the high sensitivity, compactness and stable operation at temperatures down to 12 K make our device ideal for various low vibration cryogenic experiments.

\section*{Data Availability}
The data that support the findings of this study are available from the corresponding author upon reasonable request.

\section*{Acknowledgment}
This work was supported by MEXT, JSPS Leading-edge Research Infrastructure Program, JSPS Grant-in-Aid for Specially Promoted Research 26000005, JSPS Grant-in-Aid for Scientific Research on Innovative Areas 2905: JP17H06358, JP17H06361 and JP17H06364, JSPS Core-to-Core Program A. Advanced Research Networks, JSPS Grant-in-Aid for Scientific Research (S) 17H06133, the joint research program of the Institute for Cosmic Ray Research, University of Tokyo, National Research Foundation (NRF) and Computing Infrastructure Project of KISTI-GSDC in Korea, Academia Sinica (AS), AS Grid Center (ASGC) and the Ministry of Science and Technology (MoST) in Taiwan under grants including AS-CDA-105-M06, the LIGO project, and the Virgo project. We would also like to thank KEK Machine Shop for machining the mechanical components of the accelerometer.

\end{document}